\documentclass[twocolumn]{revtex4}

\def\D#1#2{\frac{\partial #1}{\partial #2}}

\begin{document}
\title{"Dark energy" as Conformal Dynamics of Space}
%Conformal Dynamics of Space as 'Dark Energy'
\author{Dmitry E. Burlankov}%
\email{burlankov@nifti.unn.ru}
\affiliation{Physics Department, University of Nizhny Novgorod, Russia.}
\begin{abstract}
%This article presents a
An exact solution for the dynamics of conformally flat space,
which, due to the equations of dynamics, turns out to be homogeneous, is presented.
The conformal mode of space metric evolution in Global Time Theory
has negative energy density. The transfer of energy to this mode from other
modes leads to increasing homogeneity of the Universe, although the probability
of such transfer from local objects is negligibly small.
The conformal mode corresponds to 'dark energy' in observation astronomy.
\end{abstract}

\maketitle

``Dark matter'' and ``dark energy'' represent one of the fundamentally obscure issues
of modern space dynamics. Astronomical observations demonstrate that the life of
the Universe is more than just migration of stars and galaxies.
The dynamic system of the Universe has some additional degrees of freedom
known as ``dark matter'' and ``dark energy'' which make up a significant part
of the overall energy balance  \cite{Barbashov}. Modern official theory of Space and
Time - General Relativity (GR) - seems to have enough field variables,
so there is hardly any need to introduce any unclear ``dark'' substances:
the metric tensor of four-dimension space-time has ten components.
However, in the framework of GR these components cannot be considered
as a contribution to the energy balance of the Universe.

The Global Time Theory (GTT) \cite{BGL,book,GTTvGR}, which has been developed
over the past few years and is based on the physical principles that are
significantly different from General Relativity, considers three-dimensional
space as a dynamic object described by a six-component metric tensor and
a three-dimensional absolute velocity field. It means that from the point of view
of the Field Theory, space is a nine-component field. Dynamics takes place
in global time, whereas processes in non-inertial systems are described
by the mathematical technique of {\it invariant time derivations}.
%%%%%%%%%%%%%%%%%%%
Especially important for the theory is the invariant time derivative of the
metric tensor:
\begin{equation}
D_t\:\gamma _{ij}=\D{\gamma _{ij}}{t}+V_{i;j}+V_{j;i}.
\label{difmtr}
\end{equation}
%%%%%%%%%%%%%%%%%

Nine equations (according to the number of field components) are derived
from the principle of minimal action
%%%%%%%%%%%%%%%%%%%
\begin{equation}
 S=\frac{c^4}{16\,\pi\,k}\int(\mu^i_j\,\mu^j_i-(\mu^j_j)^2+R)
\sqrt{\gamma}\,d_3\,x\,dt,
\label{Lagr}
\end{equation}
where $R$ is scalar curvature of space and $\mu _{ij}$ are constructed
from invariant time derivatives of the metric tensor (\ref{difmtr}):
\begin{equation}
\mu _{ij}=\frac{1}{2\,c}D_t\,\gamma _{ij}=\frac{1}{2\,c}(\dot{\gamma}_{ij}+
V_{i;j}+V_{j;i}), \label{muij}
\end{equation}
and the absolute velocities $V^i$ are present only in the kinetic energy term.
%%%%%%%%%%%%%%%%%%%%%%%

As usual this variational technic  leads to non-trivial Hamiltonian
of space
\begin{equation}\label{HamGr}
  H=\frac{c^4}{16\pi k}\int\left(\mu^i_j\mu^j_i-(\mu^i_i)^2-
  R\right)\,\sqrt{\gamma}\,d_3x.
\end{equation}
The energy density of space can be both positive and negative.
%%%%%%%%%%%%%%%%%%%%%%%%%%%%%%%%%%%%%

If one more condition (the tenth one) is added - i.e. that energy density is equal
to zero - then these ten equations coincide with the ten equations of Einstein's
General Relativity. It means that GR solutions are contained within solutions of
GTT, which also has a multitude of other solutions with non-zero density of energy.

In cosmic dynamics only about 4\% of energy is related to matter which determines
sources in space dynamics equations. Therefore, investigating problems of pure
dynamics of space without matter can yield solutions that come quite close
to modeling the real dynamics of the Universe.

For example, the dynamics of a 3D sphere with the radius changing in time
(in a cycloid) is regarded by GTT as a problem of dynamics of space without matter.
Unlike GR, where the parameters of the cycloid are determined by the density
of matter, GTT determines these parameters by the energy of space;
in this case the dust-like matter only adds to the energy. For this reason,
problems related to ``critical density'' are non-existent in GTT, since the open
or closed construction of the Universe depends on start conditions and not on
the density of the matter.

The main dynamic variable in GTT is the six-component metric tensor of
three-dimensional space. A conformal factor can be extracted in this tensor,
all other components determining the space anisotropy. This conformal component
is the one that makes a negative contribution to kinetic energy.

However the question arises, what prevents all positive energy objects
(gravitational and electromagnetic waves, etc.) from ``falling'' to negative
energy modes?

In order to answer this question in the framework of GTT, we will consider
an extremely important and relatively simple cosmological model - namely,
the dynamics of conformally flat Universe with the following metric:
\begin{equation}\label{confmtr}
dl^2=e^{2\,u(t,x,y,z)}\,(dx^2+dy^2+dz^2).
\end{equation}

Three bond equations (obtained by variation of action by the velocity field) for this metric are:
\begin{equation}\label{lnks}
\D{}{t}\,\D{u}{x^i}=0.
\end{equation}
They lead to the separation of the time part from the space part:
\begin{equation}\label{confac}
u(t,x,y,z)=v(t)+w(x,y,z).
\end{equation}

Six dynamic equations (obtained by variation of action by metric tensor)
\begin{equation}\label{dinmtr}
(2\,\ddot{u}+3\,\dot{u}^2)\,\delta^i_j=(2\,\ddot{v}+3\,\dot{v}^2)\,\delta^i_j=G^i_j
\end{equation}
lead not only to the requirement of constant (in space coordinates) scalar curvature,
but also to homogeneity of space as a whole. In equations (\ref{dinmtr}) $G^i_j$
is Einstein's tensor of 3D space.

{\it Purely conformal dynamics requires homogeneity of space.}

Whereas, for example, in Friedman-like models space could be treated
as homogeneous for the sake of simplifying the solution, --
in pure conformal dynamics space turns out homogeneous as a result
of dynamic equations.

The above solution confirms conclusions \cite{book}  concerning the dynamics
of small perturbations of flat space (with metric tensor $\bar{\gamma}_{ij}$),
which are in many ways similar to Einstein's solutions for gravity waves in GR:
$$\gamma_{ij}=\bar{\gamma}_{ij}+h_{ij}.$$

Tensor  $h_{ij}$ is broken down into a trace-less part and conformal perturbation
of the metric:
$$h=\gamma^{ij}h_{ij}=h^i_i;\quad\chi^i_j=h^i_j-\frac{1}{3}\delta^i_j\,h.$$

The traceless part $\chi^i_j$ corresponds to Einstein's solutions of GR
and satisfies the wave equation below:
\begin{equation}\label{dinWeak}
\ddot{\chi}^i_j-\Delta\,\chi^i_j=0;\quad\chi^i_{j;i}=0;\quad \chi^i_i=0.
\end{equation}

From the GTT viewpoint, these waves have positive energy density,
which is quadratic by amplitude.

Purely conformal deformation is described by component $h$, which obeys
to a fundamentally different kind of equations:
$$\ddot{h}=0;\quad h,_j=0$$
-- where the scale changes homogeneously throughout space.

The existence of modes with positive energy and 4\% of matter distort
the purely conformal dynamics, and cause the small space heterogeneity
of conformal factor. However, in linear approximation these modes
do not interact with the conformal mode \cite{book}.

The model considered above answers a number of fundamental questions:
\begin{itemize}
  \item  If any material objects transfer their energy (or part of it)
  to the conformal mode, the homogeneity of space increases.
  \item  Since the conformal mode is homogeneous throughout the whole space,
  coefficients of interaction between various local material substances
  (e.g., electromagnetic or gravitational wave packets) and the conformal mode,
  which determine the migration of energy to the conformal mode,
  are negligibly small, and in spite of the negative energy mode,
  the Universe with positive energy exists almost independently from this mode.
\end{itemize}

The main ``contributor'' to the ``dark matter'' turns out to be velocity field.
Source \cite{grVort} (and \cite{book} as well) present the solution to the
``cosmic vortex'' problem. In this solution, the matter (in the form of stars)
does not determine the vortex dynamics, but only visualizes it in a kind of
spiral galaxy (also introducing some perturbations into the dynamics
of a pure vortex).

Author thanks N. Burlankov and S. Gubanov for discussions and
T. Bezborodova for help in translation.


\begin{thebibliography}{99}

\bibitem{Barbashov} B.M. Barbashov, V.N. Pervushin, D.V. Proscurin,
PEPAN,{\bf 34},137-189, 2003.
%Б.М.  Барбашов, В.Н. Первушин, Д.В. Проскурин.

\bibitem{BGL} D.E. Burlankov.  Procs. Int. Conf. BGL-4, p.75, N.Novgorod -- Kiev,
2004.

\bibitem{book}  D.E. Burlankov. {\it The Space Dynamics}, (in russian),
Nizhni Novgorod: NNSU, 2005.

\bibitem{GTTvGR}  Burlankov D.E. {\it arXiv: gr-qc}/0509050, 2005.

\bibitem{Einvawes} A. Einstein.  {\it Sitz. preuss. Akad. Wiss.} {\bf
1},1, 154-167, 1918.

\bibitem{grVort} D.E. Burlankov. {\it arXiv: gr-qc}/0406112, 2004.

\end{thebibliography}
\end{document}